\documentclass[twocolumn,aps,prl,showpacs,groupedaddress,superscriptaddress,nofootinbib]{revtex4-1}

\usepackage[pdftex,linkcolor=blue,citecolor=blue,urlcolor=blue,colorlinks]{hyperref}
\usepackage[dvipsnames]{xcolor}
\usepackage{graphicx,epsfig}
\usepackage{braket}
\usepackage{color}
\usepackage{cancel}
\usepackage{amsmath}
\usepackage{amsfonts}
\usepackage{fancyhdr}
\usepackage{ulem}

\begin{document}
\title{Topological edge states and Aharanov-Bohm caging with ultracold atoms carrying orbital angular momentum}
\author{G. Pelegr\'i}
\affiliation{Departament de F\'isica, Universitat Aut\`onoma de Barcelona, E-08193 Bellaterra, Spain.}
\author{A. M. Marques}
\affiliation{Department of Physics and I3N, University of Aveiro, 3810-193 Aveiro, Portugal.}
\author{R. G. Dias}
\affiliation{Department of Physics and I3N, University of Aveiro, 3810-193 Aveiro, Portugal.}
\author{A. J. Daley}
\affiliation{Department of Physics and SUPA, University of Strathclyde, Glasgow G4 0NG, United Kingdom.}
\author{J. Mompart}
\affiliation{Departament de F\'isica, Universitat Aut\`onoma de Barcelona, E-08193 Bellaterra, Spain.}
\author{V. Ahufinger}
\affiliation{Departament de F\'isica, Universitat Aut\`onoma de Barcelona, E-08193 Bellaterra, Spain.}

\begin{abstract}
We show that bosonic atoms loaded into orbital angular momentum $l=1$ states of a lattice in a diamond-chain geometry provides a flexible and simple platform for exploring a range of topological effects. This system exhibits robust edge states, and the relative phases arising naturally in the tunnelling amplitudes lead to the appearance of Aharanov-Bohm caging in the lattice. We discuss how these properties can be realised and observed in ongoing experiments.
\end{abstract}
\maketitle


Topological properties play an important role in a wide range of condensed matter systems \cite{RevTopology}. Such properties are particularly demonstrated by topological insulators \cite{RevTopIns}, where a bulk-boundary correspondence correlates the non-trivial topological indices of the bulk energy bands such as the Berry phase \cite{BerryPhase}, with the existence of topological edge states under open boundary conditions.
The importance of these concepts has led to a great deal of interest in finding clean environments in which fundamental features of the system can be observed, and phenomena arising from interactions and non-equilibrium dynamical effects can be explored. Highlights of this include the realisation of the Haldane \cite{HaldaneUltracold} and Hofstadter \cite{HofstadterUltracold1,HofstadterUltracold2} models with ultracold atoms, as well as the experimental measurement \cite{MeasureZak} of Zak's phase \cite{ZakPhase} and the detection of topological states \cite{SSHbosons1,SSHbosons2}, which complements paralell work in photonic waveguides \cite{PhotCrystal1,PhotCrystal2,PhotCrystal3,PhotCrystal4,PhotCrystal5,QWalk1,QWalk2}. There are a wide range of further theoretical proposals for observation of topological phenomena in cold atoms \cite{ReviewUltracoldTop,ProbeFermions,QWalkFermion,SPT1fermion,SPT2fermion,SPT3fermion,
QuasiPeriodic1,QuasiPeriodic2}, most of which are based around the realisation of artificial gauge fields by laser dressing \cite{ArtGaugeUltracold,HofstadterUltracold1,HofstadterUltracold2}, or periodically driving the lattice system \cite{PeriodicDrive}. 

\begin{figure}[t!]
\centering
\includegraphics[width=\linewidth]{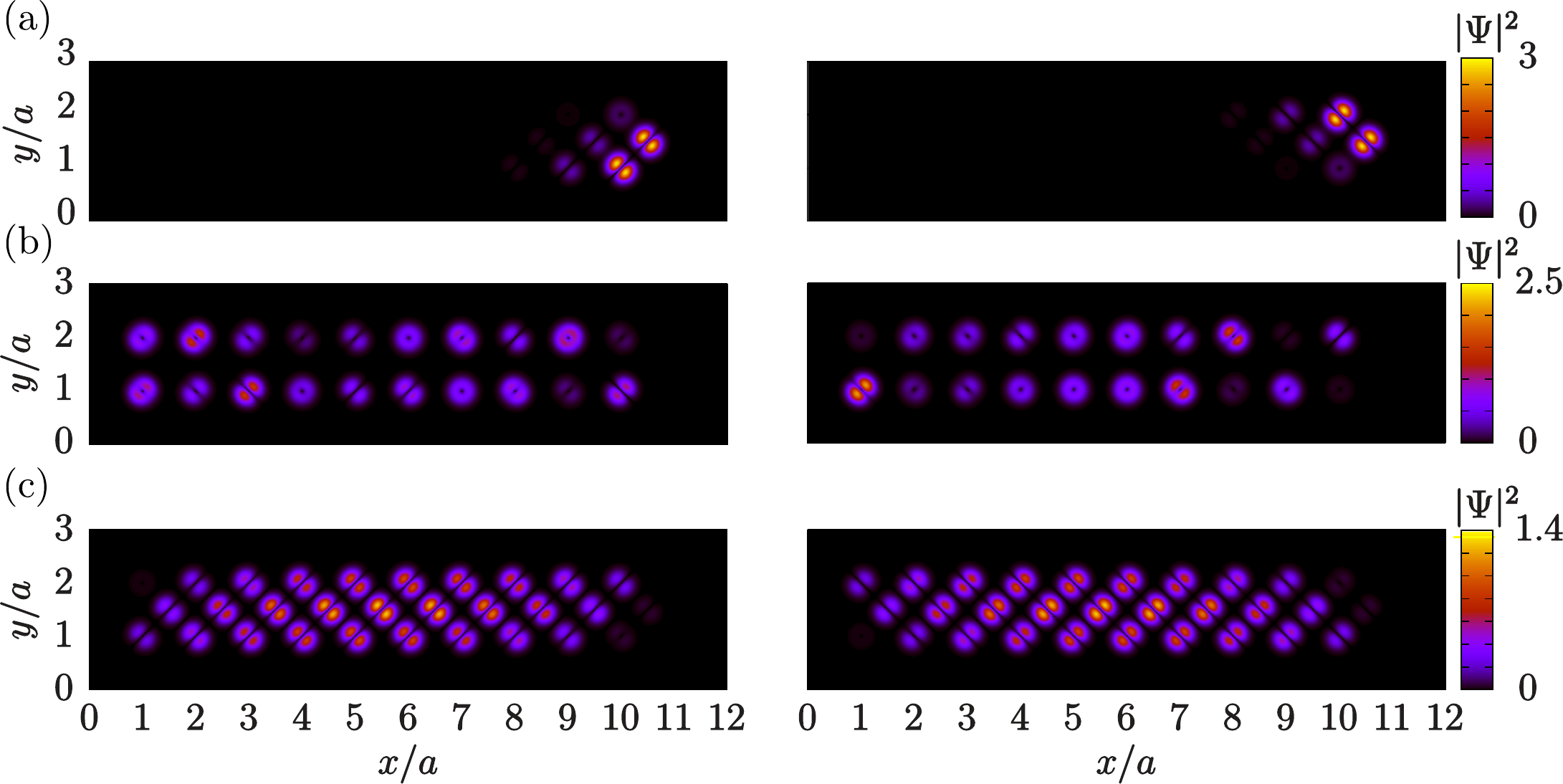}
\caption{Density profiles of numerically obtained eigenstates for a diamond chain of $N_c=10$ units cells and inter-site separation $d=6\sigma$, corresponding to $J_3/J_2=1.13$. (a) Two degenerate edge states. (b) Two states of the flat band. (c) The two degenerate ground states.}
\label{densityplots}
\end{figure}

Here, we show that an alternative and simplified approach to the realization of topologically non-trivial multi-level models in optical lattices is available by loading ultracold atoms in excited Orbital Angular Momentum (OAM) states of a 1D chain. We study a concrete example of a diamond chain to demonstrate how, due to the additional degree of freedom provided by the OAM, this model is rendered topologically non-trivial, with the respective appearance of topological states in the exact diagonalization spectrum of an open chain. The local OAM $l=1$ states are equivalent to the $p_x$ and $p_y$ orbitals in optical lattices, which have been shown to naturally display non-trivial topological properties in one-\cite{pband1D} and two-\cite{pband2D1,pband2D2} dimensional systems due to the parity of their wave functions. In the OAM $l=1$ basis as we consider here, the mechanism that yields topological properties is the appearance of relative phases in the tunnelling amplitudes. This is controllable by tuning the geometry of the lattice \cite{geometricallyinduced}.
\begin{figure}[t!]
\centering
\includegraphics[width=\linewidth]{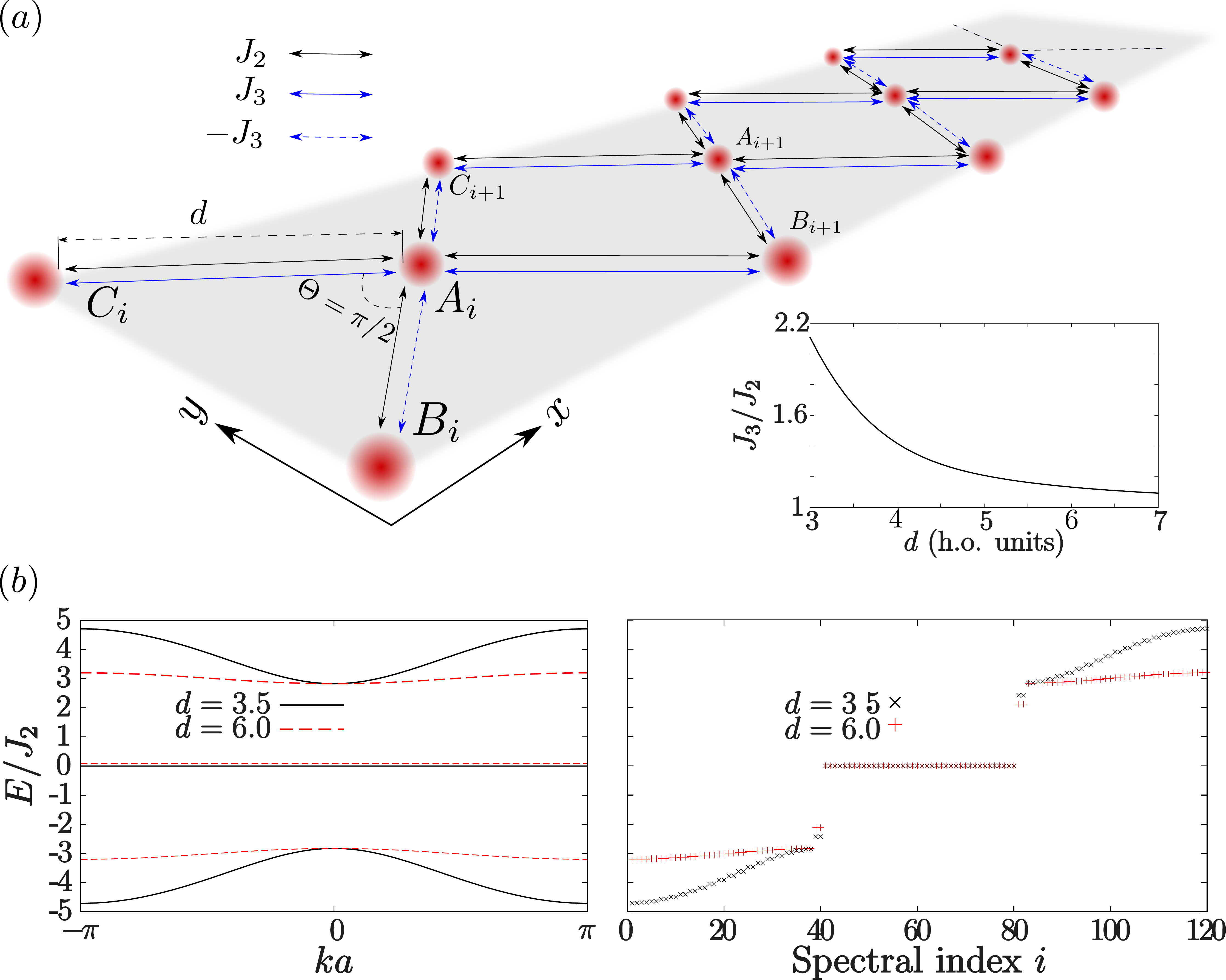}
\caption{(a) Schematic representation of the considered diamond chain. The inset shows, for harmonic oscillator traps, the dependence of the relative value of $J_2$ and $J_3$ on the inter-site separation $d$, expressed in units of $\sigma=\sqrt{\hbar/(m\omega)}$. (b) Energy spectrum of the diamond chain with OAM $l=1$ states. The left plot shows the band structure computed for $d=3.5\sigma$, corresponding to $J_3/J_2=1.67$ (black solid line) and $d=6\sigma$, corresponding to $J_3/J_2=1.13$  (red dotted line). In the right plot, the corresponding exact diagonalization spectra of a diamond chain of $N_c=20$ unit cells are shown.}
\label{PhySystem}
\end{figure}
\\
Additionally, we show how a proper tuning of the inter-site separation and the central angle can lead to destructive quantum interference that defines a spatially limited region to which specially prepared states are confined, in what is known as Aharonov-Bohm (AB) caging \cite{ABcageoriginal,ABcageoriginal2}. A distinctive advantage regarding the realization of AB caging in this model, in relation to other proposals \cite{similar2, ABcagephotonics, recentwork1, recentwork2}, is that one does not need to rely on creating synthetic gauge fields \cite{ArtGaugeFields1,ArtGaugeFields2,ArtGaugeFields3,ArtGaugeFields4} to produce the magnetic flux required for AB caging. Instead, in our OAM $l=1$ model complex phases with values controlled by the central angle appear naturally at some of the tunneling parameters \cite{geometricallyinduced, edgelike}, giving rise to an effective magnetic flux.
\\
\textit{Model.-}We consider a quasi-one-dimensional optical lattice with a diamond-chain shape. As shown in Fig.~\ref{PhySystem}(a), the unit cells of this chain, labelled with the index $i$, are formed of three sites, $A_i$, $B_i$ and $C_i$, each corresponding to a cylindrically symmetric potential of radial frequency $\omega$, and forming a triangle with central angle $\Theta=\pi/2$ and nearest-neighbour separation $d$. We assume that the lattice is composed of an integer number $N_c$ of unit cells, so that its right termination has a closed edge which, as we will show in this letter, hosts topological edge states. The chain is loaded with non-interacting ultracold atoms of mass $m$ that may occupy the two degenerate OAM $l=1$ states with positive or negative circulation localized at each site, $\braket{\vec{r}|j_i,\pm}=\psi(r_{j_i})e^{\pm i(\varphi_{j_i}-\varphi_0)}$, where $j\in\{A,B,C\}$, $(r_{j_i},\varphi_{j_i})$ are the polar coordinates with origin at the site $j_i$ and $\varphi_0$ is the phase origin. The tunneling dynamics of this type of states has been studied in detail in \cite{geometricallyinduced}. Between two neighbouring sites, there are only three independent tunneling amplitudes: $J_1$, which corresponds to the self-coupling at each site between the two OAM states with different circulations, and $J_2$ and $J_3$, which correspond to the cross-coupling between OAM states in different sites with equal or different circulations, respectively. The tunneling amplitudes between states with different OAM circulations $J_1$ and $J_3$ acquire relative phases that depend on $\varphi_0$, which is determined by $\Theta$ in the diamond-chain lattice. For $\Theta=\pi/2$, due to destructive interference between neighbouring sites with different phases in the tunneling amplitudes the self-coupling vanishes everywhere except for the sites at the left edge $B_1$ and $C_1$. Moreover, since typically $|J_1|\ll |J_2|,|J_3|$ \cite{geometricallyinduced}, in this letter we neglect the self-coupling term at these two sites and leave a study of its consequences for Ref.~\cite{otherpaper}. Choosing $\varphi_0$ to point along the direction of the line that connects the sites $C_i$, $A_i$ and $B_{i+1}$ and assuming coupling only between nearest-neighbouring sites due to the rapid decay of the tunneling amplitudes as the inter-site separation increases \cite{otherpaper}, the non-interacting Hamiltonian of the system takes the form ($\hbar\equiv 1$)
\begin{align}
\hat{H}&=J_2\sum_{i=1}^{N_c}\sum_{\alpha=\pm}\left[\hat{a}_{\alpha}^{i\dagger}(\hat{b}_{\alpha}^{i}+\hat{b}_{\alpha}^{i+1}+\hat{c}_{\alpha}^{i}+\hat{c}_{\alpha}^{i+1})\right]+\text{h.c.}\nonumber\\
&+J_3\sum_{i=1}^{N_c}\sum_{\alpha=\pm}\left[\hat{a}_{\alpha}^{i\dagger}(e^{i\pi}\hat{b}_{-{\alpha}}^{i}+\hat{b}_{-{\alpha}}^{i+1}
+\hat{c}_{-{\alpha}}^{i}+e^{i\pi}\hat{c}_{-{\alpha}}^{i+1})\right]+\text{h.c.},
\label{Hoperators}
\end{align}
In our convention for $\varphi_0$, we see that a $\pi$ phase is acquired in tunnelling $B_i\leftrightarrow A_i \leftrightarrow C_{i+1}$ for a central angle $\Theta=\pi/2$. As shown in the inset of Fig.~\ref{PhySystem}(a), the relative value of $J_2$ and $J_3$ depends on the inter-site separation $d$, starting at $J_3/J_2\approx 2.2$ for $d=3\sigma$ and tending rapidly and asymptotically to $J_3/J_2=1$ as $d$ increases. By rearranging the diamond lattice with two states per site described by \eqref{Hoperators} as a one-dimensional chain with long-range couplings, it can be shown that the model possesses inversion symmetry, leading to a quantization to $0$ or $\pi$ ${(\text{mod } 2\pi)}$ of the Zak's phases \cite{ZakPhase}. Additionally, since the model is bipartite it has chiral symmetry defined as $\Gamma\hat{H}\Gamma=-\hat{H}$, which entails that the energy spectrum is symmetric around 0.
\\
By applying a series of exact mappings, we shall demonstrate that these symmetries are accompanied by the presence of topologically protected states localized at the right edge of the chain.
Under periodic boundary conditions, the bulk Hamiltonian corresponding to the Fourier transform of \eqref{Hoperators} yields six energy bands after diagonalization. Their dispersion relations appear in three degenerate pairs ${E(k)=0,\pm 2\sqrt{(J_2^2+J_3^2)+\cos(ka)(J_2^2-J_3^2)}}$, where $a=\sqrt{2}d$ is the lattice constant. The band structure presents a gap of size $2\sqrt{2}J_2$ (for $J_3>J_2$) or $2\sqrt{2}J_3$ (for $J_3<J_2$) and, in the $J_2=J_3$ limit, all bands become flat. As shown in the energy spectrum of Fig.~\ref{PhySystem}(b), for an experimentally feasible inter-site separation of $d=6\sigma$ one is already very close to this all-flat limit. In the case of open boundary conditions, exact diagonalization performed for a chain with $N_c=20$ unit cells, shown in Fig.~\ref{PhySystem} (b), reveals the presence of four in-gap states localized at the right edge of the chain. These in-gap states persist provided both $J_2$ and $J_3$ are non-zero, that is, as long as there is an energy gap.
\begin{figure}[t!]
\centering
\includegraphics[width=\linewidth]{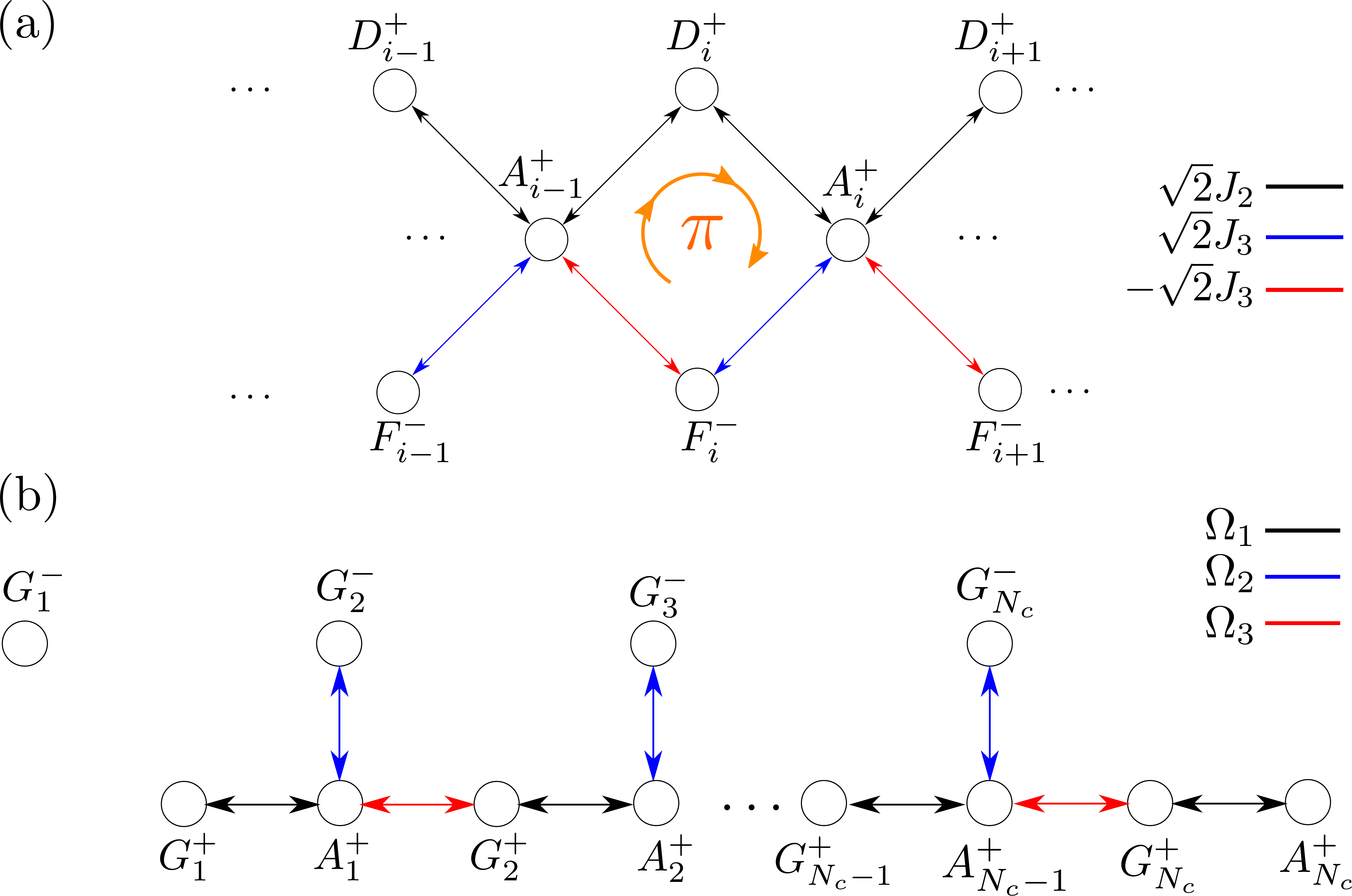}
\caption{Schematic representation of the tight-binding models obtained through the different mappings. (a) $H^+$ diamond chain obtained after applying the basis rotation $\{\ket{B_i,\pm},\ket{C_i,\pm}\}\rightarrow \{\ket{F_i,\pm},\ket{D_i,\pm}\}$ to the original OAM $l=1$ diamond chain. (b) Modified SSH model obtained after applying the basis rotation $\{\ket{D_i,+},\ket{F_i,-}\}\rightarrow \{\ket{G_i,\pm}\}$ to the $H^+$ chain. $N_c$ is the total number of unit cells.}
\label{mappings}
\end{figure}
\\
\textit{Analytical mappings.-}The two-fold degeneracy of the spectrum and the presence of gaps in the band structure can be understood by performing {a rotation into a basis of symmetric and antisymmetric states}, $\ket{D_i,\pm}=\frac{1}{\sqrt{2}}(\ket{C_i,+}\pm\ket{B_i,+})$, $\ket{F_i,\pm}=\frac{1}{\sqrt{2}}(\ket{C_i,-}\pm\ket{B_i,-})$. {This rotation} decouples the diamond chain with six states per unit cell \eqref{Hoperators} into two independent and identical diamond chains, $H^+$ and $H^-$, which have three states per unit cell, see Fig.~\ref{mappings}(a). These two chains are described by the Hamiltonians
\begin{subequations}
\begin{align}
\hat{H}^{+}&=\sum_{i=1}^{N_c} \hat{a}_+^{i\dagger}[\sqrt{2}J_2(\hat{d}_+^{i}+\hat{d}_+^{i+1})+\sqrt{2}J_3(\hat{f}_-^{i}-\hat{f}_-^{i+1})]+\text{h.c.}\\
\hat{H}^{-}&=\sum_{i=1}^{N_c} \hat{a}_-^{i\dagger}[\sqrt{2}J_2(\hat{f}_+^{i}+\hat{f}_+^{i+1})+\sqrt{2}J_3(\hat{d}_-^{i}-\hat{d}_-^{i+1})]+\text{h.c.}
\end{align}
\label{HamsDecoupledfirstmapping}
\end{subequations}
The minus sign in one of the couplings can be associated with a net $\pi$ flux through the plaquettes of these diamond chains \cite{otherpaper}, which explains the gap in the band structure \cite{fermionsdiamondchain}. 
\\
Furthermore, the existence of in-gap edge states and the flattening of the bands in the $J_2=J_3$ limit can be understood by means of a second basis rotation for each of the two subchains $H^+$ and $H^-$. For the $H^+$ chain, this basis rotation is given by $\ket{G_i,+}=\frac{1}{\sqrt{J_2^2+J_3^2}}(J_2\ket{D_i,+}+J_3\ket{F_i,-})$ and $\ket{G_i,-}=\frac{1}{\sqrt{J_2^2+J_3^2}}(J_3\ket{D_i,+}-J_2\ket{F_i,-})$. An equivalent rotation can be defined for the $H^-$ chain by replacing $F\leftrightarrow D$. After applying this transformation, the $H^+$ diamond chain is mapped into a modified Su-Schrieffer-Heeger (SSH) model \cite{SSHoriginal} with an extra dangling state per unit cell, as shown in Fig.~\ref{mappings}(b), which is described by the Hamiltonian
\begin{equation}
\hat{H}^{+}_{SSH}=\sum_{i=1}^{N_c}\hat{a}_+^{i\dagger}(\Omega_1\hat{g}_+^{i}+\Omega_2\hat{g}_-^{i+1}+\Omega_3\hat{g}_+^{i+1})+\text{h.c.},
\label{HamSSH}
\end{equation}
where $\Omega_1=\sqrt{2}\sqrt{J_2^2+J_3^2}$ and $\Omega_3=\frac{\sqrt{2}(J_2^2-J_3^2)}{\sqrt{J_2^2+J_3^2}}$ are the strong and weak horizontal couplings and $\Omega_2=\frac{2\sqrt{2}J_2J_3}{\sqrt{J_2^2+J_3^2}}$ is the coupling strength of the dangling state of each unit cell $i$ to the central state $\ket{A_i,+}$. From eq. \eqref{HamSSH}, a compact expression for the zero-energy flat band states $\hat{H}^{+}_{SSH}\ket{0}^+_i=0$ can be derived
\begin{equation}
\ket{0}_i^+=\frac{1}{\sqrt{C}}\left(\frac{\Omega_3}{\Omega_2}\ket{G_i,-}-\ket{G_i,+}+\frac{\Omega_1}{\Omega_2}\ket{G_{i+1},-}\right).
\label{zeroenerg}
\end{equation}
In the original basis $\{\ket{j_i,\pm}\}$, the most localized forms of the zero-energy states \eqref{zeroenerg} span the four sites surrounding the central site $A_i$, with no contribution from the states at this site. Additionally, $\ket{G_1,-}$ is a completely decoupled zero-energy state localized at the left edge of the chain, see Fig.~\ref{mappings} (b). In the $J_2=J_3$ limit, we have $\Omega_3=0$ and the bulk of the chain can be decomposed into isolated trimers $\{\ket{G_i,+},\ket{A_i,+},\ket{G_{i+1},-}\}$ with equal internal couplings $\Omega_1=\Omega_2=2J_2$. The eigenstates of these trimers with a component of the state $\ket{A_i,+}$ are the top and bottom flat-band states $\ket{E\pm}_i^+$
\begin{align}
\ket{E\pm}_i^+ &=\frac{1}{2}\left(\ket{G_i,+}\pm\sqrt{2}\ket{A_i,+}+\ket{G_{i+1},-}\right);\nonumber\\ &\hat{H}^+_{SSH}|_{\Omega_3=0}\ket{E\pm}_i^+=\pm 2\sqrt{2}J_2\ket{E\pm}_i^+.
\label{flatbandstatesomega3zero}
\end{align} 
However, at the right edge of the chain there is a dimer formed by the states $\ket{G_{N_c},+}$ and $\ket{A_{N_c},+}$, whose eigenstates $\ket{\text{Edge},\pm}^+$ have the following expressions and energies
\begin{align}
&\ket{\text{Edge}\pm}^+=\frac{1}{\sqrt{2}}(\ket{G_{N_c},+}\pm\ket{A_{N_c},+});\nonumber\\
&\hat{H}^+_{SSH}|_{\Omega_3=0}\ket{\text{Edge}\pm}^+=\pm 2J_2\ket{\text{Edge}\pm}^+.
\label{edgestates}
\end{align}
\\
As shown in Fig.~\ref{PhySystem} (b), in-gap states appear also for general values of the couplings $J_2\neq J_3$. In this scenario, these states are strongly localized at the right edge of the chain, exhibiting an exponentially decaying tail to the bulk, which widens as one deviates from the $J_2=J_3$ limit \citep{otherpaper}. In order to tell whether this robustness is due to topological effects, one should compute the Zak's phase for each band, which are the relevant quantities to topologically characterize one-dimensional models \cite{ZakPhase}. However, the computation of the Zak's phases is not straightforward in our system. In the original OAM $l=1$ model \eqref{Hoperators} the degeneracy of the bands means that their Zak's phases are ill-defined. On the other hand, each of the decoupled chains of the two successive mappings given by the Hamiltonians in \eqref{HamsDecoupledfirstmapping} and \eqref{HamSSH}, respectively, does not have inversion symmetry, so that the Zak's phase can yield non-quantised values. In order to circumvent these limitations a third mapping can be introduced, through a basis rotation of (3) (see \cite{otherpaper} for details), wherein inversion symmetry and, therefore, a quantized Zak's phase for each band, is recovered. Under this third mapping the system becomes a diamond chain with alternating tunneling amplitudes, whose non-trivial topological nature of the gaps where the edge states lie is explicitly shown in \cite{topologicalSSHring,ZakGeneralization}. A striking feature of the topology of this model, directly carried over to the original OAM $l=1$ model, is that there is no topological transition across the gap closing point, as can be seen by fixing either $J_2$ or $J_3$ and varying the other across zero. 
\\
\textit{Numerical results.-}To illustrate these results, in Fig.~\ref{densityplots} we show the numerical density plots of the different types of states that can be found in a diamond chain of $N_c=10$ unit cells and a separation between nearest-neighbour sites $d=6\sigma$, corresponding to $J_3/J_2=1.13$. In (a), two degenerate edge states are shown, evidencing their strong localization at the right end of the chain. In (b), two examples of zero-energy states, which have no population at the central ($A$) sites of the chain, are displayed. These contain components of many maximally localized states \eqref{zeroenerg} and, in the case of the state at the right panel, also of the zero-energy decoupled mode localized at the left edge,$\ket{G_1,-}$, see Fig.~\ref{mappings} (b). Finally, (c) shows a the two degenerate ground states of the system. In (a) and (c), the states at the left and right panels have different orientations of the nodal lines due to the fact that they belong to the two different subchains $H^+$ and $H^-$. 

\begin{figure}[t!]
\centering
\includegraphics[width=\linewidth]{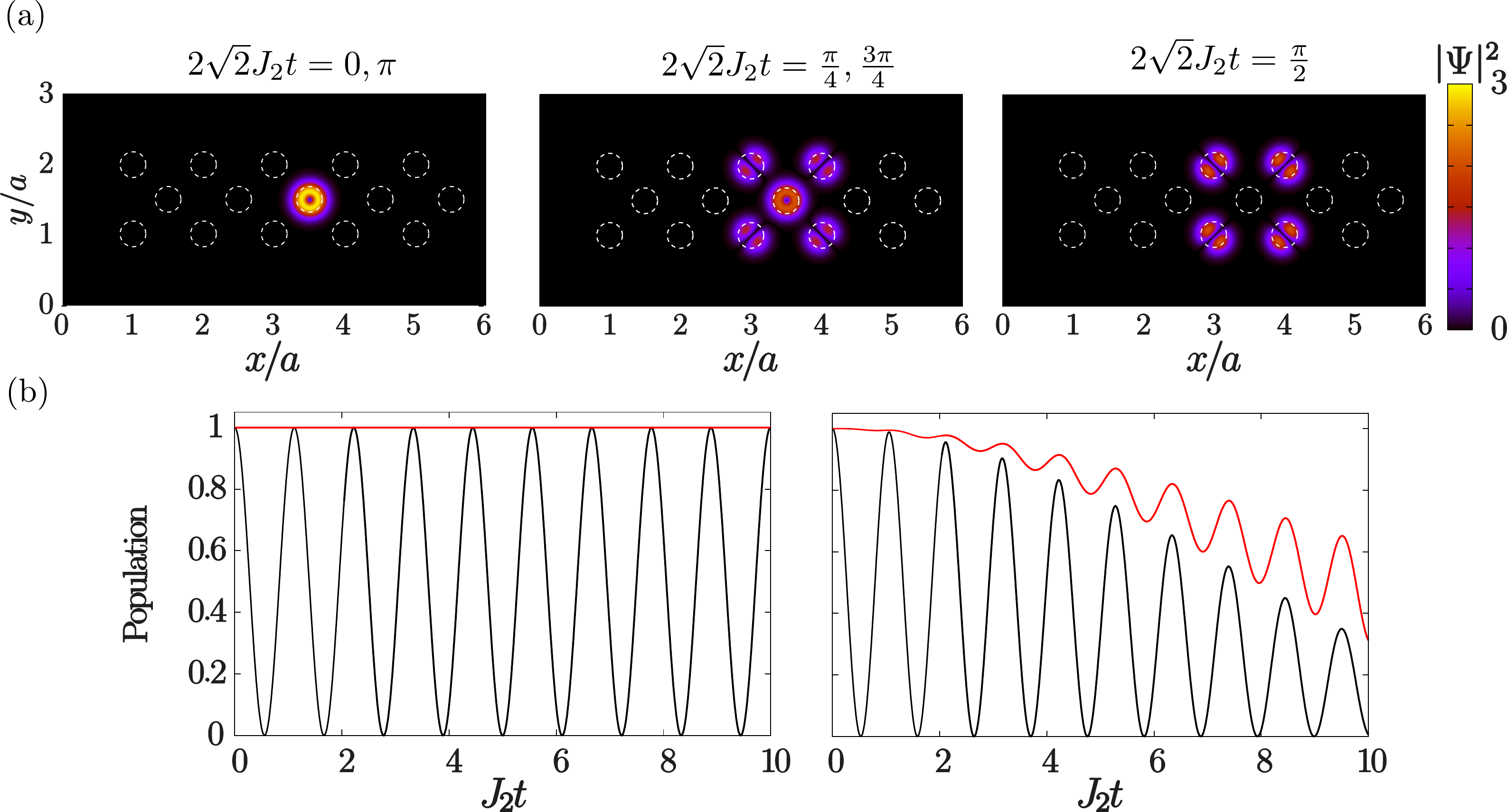}
\caption{AB caging in a diamond chain of $N_c=5$ unit cells. (a) Snapshots at different times of the density profiles corresponding to the time evolution of a wave packet initially prepared in the state $\ket{A_3,+}$ in the perfect caging limit $J_2/J_3=1$. (b) Time evolution of the population of the same initial state as in (a) (black solid lines) and the total population of the cage (red slid lines) for $J_3/J_2=1$ (left panel) and 1.1 (right panel).}
\label{AharanovBohm}
\end{figure}

\textit{Aharanov-Bohm caging.-}Finally, we show that in the $J_2=J_3$ limit the system can exhibit AB caging. In this limit, from the relations \eqref{flatbandstatesomega3zero} and the equivalent ones for the $H^-$ chain, we can express the states $\ket{A_i,\pm}$ in terms of flat-band states that occupy solely the four sites surrounding $A_i$, i.e., $B_i$, $B_{i+1}$, $C_i$ and $C_{i+1}$. Therefore, an initial state prepared in an arbitrary superposition of the $\ket{A_i,+}$ and $\ket{A_i,-}$ states will oscillate coherently to its four neighbouring sites with a frequency $\omega=2\sqrt{2}J_2$ (given by the absolute value of the energies of the top/bottom flat-band states) and, therefore, never leave the cage formed by the unit cells $i$ and $i+1$. In Fig.~\ref{AharanovBohm} (a), we show some snapshots of the time evolution of a wave packet prepared initially in the state $\ket{A_3,+}$ of a diamond chain with $N_c=5$ unit cells. In a real experiment, the condition $J_2=J_3$ would never be exactly fulfilled, but for sufficiently close values the AB caging would persist for a significant amount of time. In Fig.~\ref{AharanovBohm} (b), we plot, for the same initial state as in (a), the time evolution of the population of the state $\ket{A_3,+}$ and the total sum of the states forming the cage for $J_3/J_2=1,1.1$ (left and right panels, respectively). While perfect caging only occurs for $J_3/J_2=1$, for $J_3/J_2=1.1$ we observe that approximately $40\%$ of the population remains on the cage after a time $J_2t=10$.  
\\
\textit{Experimental implementation.-} An optical diamond chain could be implemented by using two pairs of counter-propagating lasers at $\pm 45$ degrees with respect to the $x-$axis in a quasi 1D-cigar shape geometry \cite{FlatBandLattice}. To load atoms in the OAM $l=1$ manifold of local sites of the lattice, which are combinations of the $p-$band orbitals of the form $p_x\pm ip_y$, three different approaches could be used:  to adiabatically modify the trapping potentials such that atoms are transferred from the ground to the $p-$band of an adjacent well via resonant tunnelling \cite{SuperfluidpBand}, to combine lattice shaking with shortcuts to adiabaticity to promote the atoms to the $p-$band \cite{HigherOrbital2}, and, finally, to directly transfer OAM from a light beam to the trapped atoms \cite{OptOAMAtoms}. Once the atoms are loaded into the $p$-band, loss of population can be induced by collisions that transfer one atom to the lowest band, and one to a higher band. This is strongly suppressed for dilute samples and weak interactions, and in deep lattices where bandwidths are small anharmonicity ensures that the process is not resonant. To detect the atomic distribution and thereby AB caging and edge states with single-site resolution in the diamond chain, the quantum gas microscope technique \cite{QGasMicro1,QGasMicro2} could be used. This technique has recently been applied to observe topological states of ultracold bosonic atoms in optical lattices \cite{TopologicalOptLat1,TopologicalOptLat2}. On the other hand, edge states have been observed with an atomic Bose gas in the quantum Hall regime \cite{edgeHall} making use of the synthetic dimensions provided by the internal degrees of freedom.
\\
\textit{Conclusions.-} In summary, OAM provides phases in the tunneling amplitudes, which can be tuned by modifying the geometry. For realistic experimental parameters, this can be used to directly observe topological edge states and AB caging in a diamond chain.  This could form the basis for future studies of interacting particles, and also a broad range of scenarios of out of equilibrium dynamics in topological lattices.
\section*{Acknowledgements}
GP, JM and VA gratefully acknowledge financial support
through  the  Spanish  Ministry  of  Science  and  Innovation (MINECO) (Contract No. FIS2014-57460P) and the Catalan Government (Contract No.  SGR2014-1639). GP acknowledges inancial support from MINECO through Grant No. BES-2015-073772. AMM acknowledges financial support from the Portuguese Institute for Nanostructures, Nanomodelling and Nanofabrication (i3N) through the grant BI/UI96/6376/2018. Work at the University of Strathclyde was supported by the EPSRC Programme Grant
DesOEQ (EP/P009565/1). We thank Alessio Celi, Alexandre Dauphin, Anton Buyskikh and Stuart Flannigan for helpful discussions.


\begin{thebibliography}{99}
\bibitem{RevTopology} X.-L. Qi and Shou-Cheng Zhang, \textit{Rev. Mod. Phys.} \textbf{83}, 1057 (2011).

\bibitem{RevTopIns} M. Z. Hasan and C. L. Kane, \textit{Rev. Mod. Phys.} \textbf{82}, 3045 (2010).

\bibitem{BerryPhase} M. V. Berry, \textit{Proc. R. Soc. Lond. A Math. Phys. Sci.} \textbf{392}, 1802 (1984).


\bibitem{HaldaneUltracold} G. Jotzu, M. Messer, R. Desbuquois, M. Lebrat, T. Uehlinger, D. Greif, and T. Esslinger, \textit{Nature} \textbf{515}, 237 (2014).

\bibitem{HofstadterUltracold1} M. Aidelsburger, M. Atala, M. Lohse, J. T. Barreiro, B. Paredes, and I. Bloch, \textit{Phys. Rev. Lett.} \textbf{111}, 185301 (2013).

\bibitem{HofstadterUltracold2} H. Miyake, G. A. Siviloglou, C. J. Kennedy, W. C. Burton, and W. Ketterle, \textit{Phys. Rev. Lett.} \textbf{111}, 185302 (2013).

\bibitem{MeasureZak} M. Atala, M. Aidelsburger, J. T. Barreiro, D. Abanin, T. Kitagawa, E. Demler, and I. Bloch, \textit{Nature Physics} \textbf{9}, 795 (2013).

\bibitem{ZakPhase} J. Zak, \textit{Phys. Rev. Lett.} \textbf{62}, 2747 (1989).

\bibitem{SSHbosons1} M. Leder, C. Grossert, L. Sitta, M. Genske, A. Rosch, and M. Weitz, \textit{Nature Communications} \textbf{7}, 13112 (2016).

\bibitem{SSHbosons2} E. J. Meier, F. A. An, and B. Gadway \textit{Nature Communications} \textbf{7}, 13986 (2016).

\bibitem{PhotCrystal1} Y. E. Kraus, Y. Lahini, Z. Ringel, M. Verbin, and O. Zilberberg
\textit{Phys. Rev. Lett.} \textbf{109}, 106402 (2012).

\bibitem{PhotCrystal2} M. Verbin, O. Zilberberg, Y. E. Kraus, Y. Lahini, and Y. Silberberg
\textit{Phys. Rev. Lett.} \textbf{110}, 076403 (2013)

\bibitem{PhotCrystal3} M. Hafezi, S. Mittal, J. Fan, A. Migdall, and J. M. Taylor \textit{Nature Photonics} \textbf{7}, 1001 (2013).

\bibitem{PhotCrystal4} P. St-Jean, V. Goblot, E. Galopin, A. Lemaître, T. Ozawa, L. Le Gratiet, I. Sagnes, J. Bloch, and A. Amo, \textit{Nature Photonics} \textbf{11}, 651 (2017).

\bibitem{PhotCrystal5} S. Weimann, M. Kremer, Y. Plotnik, Y. Lumer, S. Nolte, K. G. Makris, M. Segev, M. C. Rechtsman, A. Szameit \textit{Nature Materials} \textbf{16}, 433 (2017).

\bibitem{QWalk1} T. Kitagawa, M. A. Broome, A. Fedrizzi, M. S. Rudner, E. Berg, I. Kassal, A. Aspuru-Guzik, E. Demler, and A. G. White, \textit{Nature Communications} \textbf{3}, 882 (2012).

\bibitem{QWalk2} F. Cardano, A. D'Errico, A. Dauphin, M. Maffei, B. Piccirillo, C. de Lisio, G. De Filippis, V. Cataudella, E. Santamato, L. Marrucci, M. Lewenstein, and P. Massignan, \textit{Nature Communications} \textbf{8}, 15516 (2017). 

\bibitem{ReviewUltracoldTop} N. Goldman, J. C. Budich, and P. Zoller, \textit{Nature Phyiscs} \textbf{12}, 639 (2016).

\bibitem{ProbeFermions}  M. Metcalf, C.-Y. Lai, K. Wright, and C.-C. Chien, \textit{Europhysics Letters} \textbf{118}, 56004 (2017). 

\bibitem{QWalkFermion} S. Mugel, A. Celi, P. Massignan, J. K. Asb\'oth, M. Lewenstein, and C. Lobo,
\textit{Phys. Rev. A} \textbf{94}, 023631 (2016).

\bibitem{SPT1fermion} H. Nonne, M. Moliner, S. Capponi, P. Lecheminant and K. Totsuka, \textit{Europhysics Letters} \textbf{102} 37008 (2013).

\bibitem{SPT2fermion} X.-J. Liu, Z.-X. Liu, and M. Cheng, \textit{Phys. Rev. Lett.} \textbf{110}, 076401 (2013).

\bibitem{SPT3fermion} M. Nakagawa and N. Kawakami, \textit{Phys. Rev. B} \textbf{96}, 155133 (2017).

\bibitem{QuasiPeriodic1} F. Matsuda, M. Tezuka, and N. Kawakami, \textit{J. Phys. Soc. Jpn.} \textbf{83}, 083707 (2014).

\bibitem{QuasiPeriodic2} X. Deng and L. Santos, \textit{Phys. Rev. A} \textbf{89}, 033632 (2014).

\bibitem{ArtGaugeUltracold} D. Jaksch and P. Zoller, \textit{New J. Phys.} \textbf{5}, 56 (2003).

\bibitem{PeriodicDrive} K. Jim\'enez-Garc\'ia, L. J. LeBlanc, R. A. Williams, M. C. Beeler, A. R. Perry, and I. B. Spielman, \textit{Phys. Rev. Lett.} \textbf{108}, 225303 (2012).

\bibitem{pband1D} X. Li, E. Zhao, and W. Vincent Liu, \textit{Nature Communications} \textbf{4}, 1523 (2013).

\bibitem{pband2D1} K. Sun, W. V. Liu, A. Hemmerich, and S. Das Sarma, \textit{Nature Physics} \textbf{8}, 67 (2012).

\bibitem{pband2D2} Z.-F. Xu, L. You, A. Hemmerich, and W. V. Liu, \textit{Phys. Rev. Lett.} \textbf{117}, 085301 (2016).

\bibitem{geometricallyinduced} J. Polo, J. Mompart, and V. Ahufinger, \textit{Phys. Rev. A} \textbf{93}, 033613 (2016). 








\bibitem{ABcageoriginal} J. Vidal, R. Mosseri, and B. Dou\c{c}ot, \textit{Phys. Rev. Lett.} \textbf{81}, 5888 (1998).

\bibitem{ABcageoriginal2} B. Dou\c{c}ot and J. Vidal, \textit{Phys. Rev. Lett.} \textbf{88}, 227005 (2002).

\bibitem{similar2} J. J\"unemann, A. Piga, S.-J. Ran, M. Lewenstein, M. Rizzi, and A. Bermudez
\textit{Phys. Rev. X} \textbf{7}, 031057 (2017).

\bibitem{ABcagephotonics} S. Longhi, \textit{Opt. Lett.} \textbf{39}, 5892 (2014).

\bibitem{recentwork1} S. Mukherjee, M. Di Liberto, P. \"Ohberg, R. R. Thomson, and N. Goldman, arXiv: 1805.03564 [physics.optics].

\bibitem{recentwork2} M. Kremer, I. Petrides, E. Meyer, M. Heinrich, O. Zilberberg, and A. Szameit, arXiv: 1805.05209 [cond-mat.mes-hall].

\bibitem{ArtGaugeFields1}  J. Dalibard, F. Gerbier, G. Juzeli\={u}nas, and P. \"Ohberg, \textit{Rev. Mod. Phys.} \textbf{83}, 1523 (2011).

\bibitem{ArtGaugeFields2} N. Goldman, G. Juzeli\={u}nas, P. \"Ohberg, and I. B. Spielman, \textit{Rep. Prog. Phys.} \textbf{77}, 126401 (2014).

\bibitem{ArtGaugeFields3} N. Goldman and J. Dalibard, \textit{Phys. Rev. X} \textbf{4}, 031027 (2014).

\bibitem{ArtGaugeFields4} M. Aidelsburger, S. Nascimbene, and N. Goldman arXiv:1710.00851 [cond-mat.mes-hall]. 

\bibitem{edgelike} G. Pelegr\'i, J. Polo, A. Turpin, M. Lewenstein, J. Mompart, and V. Ahufinger
\textit{Phys. Rev. A} \textbf{95}, 013614 (2017).

\bibitem{otherpaper} G. Pelegr\'i, A. M. Marques, R. G. Dias, A. J. Daley, V. Ahufinger, and J. Mompart, arxiv: 1807.07533

\bibitem{fermionsdiamondchain} A. A. Lopes and R. G. Dias, \textit{Phys. Rev. B} \textbf{84}, 085124 (2011).

\bibitem{SSHoriginal} W. P. Su, J. R. Schrieffer, and A. J. Heeger \textit{Phys. Rev. Lett.} \textbf{42}, 1698 (1979).

\bibitem{topologicalSSHring} A. M. Marques and R. G. Dias, \textit{J. Phys.: Condens. Matter} \textbf{30}, 305601 (2018).

\bibitem{ZakGeneralization} A. M. Marques and R. G. Dias, arXiv:1707.06162 [cond-mat.str-el].

\bibitem{FlatBandLattice} M. Hyrk\"as, V. Apaja, and M. Manninen, \textit{Phys. Rev. A} \textbf{87}, 023614(2013).

\bibitem{HigherOrbital1} X. Li and W. V. Liu, \textit{Rep. Prog. Phys.} \textbf{79} 116401 (2016).

\bibitem{HigherOrbital2} A. Kiely, A. Benseny, T. Busch and A. Ruschhaupt, \textit{J. Phys. B} \textbf{49}, 215003 (2016).

\bibitem{HigherOrbital3} T. Kock, C. Hippler, A. Ewerbeck, and A. Hemmerich, \textit{J. Phys. B: At. Mol. Opt. Phys.} \textbf{49}, 042001 (2016).

\bibitem{OptOAMAtoms} S. Franke-Arnold, \textit{Phil. Trans. R. Soc. A} \textbf{375} 2087 (2017).

\bibitem{SuperfluidpBand} G. Wirth, M. \"Olschl\"ager, and A. Hemmerich, \textit{Nat. Phys.} \textbf{7}, 147 (2011).

\bibitem{QGasMicro1} W. S. Bakr, J. I. Gillen, A. Peng, S. F\"olling, and M. Greiner, \textit{Nature} \textbf{462}, 74 (2009).

\bibitem{QGasMicro2} J. F. Sherson, C. Weitenberg, M. Endres, M. Cheneau, I. Bloch, and S. Kuhr, \textit{Nature} \textbf{467}, 68 (2010).

\bibitem{TopologicalOptLat1} M. Aidelsburger, M. Lohse, C. Schweizer, M. Atala, J. T. Barreiro, S. Nascimb\`ene, N. R. Cooper, I. Bloch, and N. Goldman, \textit{Nat. Phys.} \textbf{11}, 162 (2015).

\bibitem{TopologicalOptLat2} M. E. Tai, A. Lukin, M. Rispoli, R. Schittko, T. Menke, D. Borgnia, P. M. Preiss, F. Grusdt, A. M. Kaufman, and M. Greiner, \textit{Nature} \textbf{546}, 519 (2017).

\bibitem{edgeHall} B. K. Stuhl, H.-I. Lu, L. M. Aycock, D. Genkina, and I. B. Spielman, \textit{Science} \textbf{349}, 1514 (2015).

 
\end{thebibliography}
\end{document}